\documentclass[sigconf,screen]{acmart}

\copyrightyear{2026}
\acmYear{2026}
\setcopyright{cc}
\setcctype{by}
\acmConference[ASE '26]{Proceedings of the 41st IEEE/ACM International Conference on Automated Software Engineering}{October 12--16, 2026}{Munich, Germany}
\acmBooktitle{Proceedings of the 41st IEEE/ACM International Conference on Automated Software Engineering (ASE '26), October 12--16, 2026, Munich, Germany}
\acmDOI{10.1145/3832783.3834531}
\acmISBN{979-8-4007-2882-2/2026/10}
\acmSubmissionID{ase26ind-p408-p}
\received{2026-05-01}
\received[accepted]{2026-07-01}

\usepackage{booktabs}
\usepackage{listings}
\usepackage{xcolor}
\usepackage{url}
\usepackage{hyperref}
\usepackage{graphicx}
\usepackage{amsmath}
\usepackage{tabularx}
\usepackage{microtype}
\usepackage{siunitx}
\usepackage{float}
\usepackage{balance}

\DeclareSIUnit{\percentagepoint}{pp}
\DeclareSIUnit{\loc}{LOC}
\AtBeginEnvironment{table}{%
  \setlength{\abovecaptionskip}{2pt}%
  \setlength{\belowcaptionskip}{2pt}%
  \setlength{\tabcolsep}{3pt}%
  \scriptsize
}
\newcommand{\code}[1]{\texttt{#1}}
\definecolor{KeyIdeaBlue}{HTML}{006BFF}
\definecolor{AlgPanelBg}{HTML}{F8FAFC}
\definecolor{AlgPanelBorder}{HTML}{2563EB}
\definecolor{AlgDecisionBg}{HTML}{FFF7ED}
\definecolor{AlgDecisionBorder}{HTML}{EA580C}
\makeatletter
\floatstyle{ruled}
\newfloat{algorithm}{t}{loa}
\floatname{algorithm}{Algorithm}
\newcounter{AlgoLine}
\newlength{\AlgoIndent}
\newcommand{\AlgoReset}{\setcounter{AlgoLine}{0}\setlength{\AlgoIndent}{0pt}}
\newcommand{\AlgoTextStyle}{\scriptsize\setlength{\baselineskip}{6.8pt}}
\newcommand{\AlgoLine}[1]{%
  \refstepcounter{AlgoLine}%
  \par\noindent
  \makebox[1.8em][r]{\scriptsize\arabic{AlgoLine}:}%
  \hspace{0.45em}\hspace*{\AlgoIndent}%
  \begin{minipage}[t]{\dimexpr\linewidth-2.25em-\AlgoIndent\relax}%
    \raggedright\AlgoTextStyle #1%
  \end{minipage}\par
}
\newcommand{\AlgoMeta}[2]{%
  \par\noindent\hspace*{2.25em}%
  \begin{minipage}[t]{\dimexpr\linewidth-2.25em\relax}%
    \raggedright\AlgoTextStyle\textbf{#1:} #2%
  \end{minipage}\par
}
\AtBeginEnvironment{algorithm}{\AlgoReset\setlength{\parskip}{0pt}\setlength{\abovecaptionskip}{1pt}\setlength{\belowcaptionskip}{1pt}}
\newcommand{\AlgoStmt}[1]{\AlgoLine{#1}}
\newcommand{\AlgoNote}[1]{\AlgoLine{$\triangleright$ \textit{#1}}}
\newcommand{\AlgoInput}[1]{\AlgoMeta{Input}{#1}}
\newcommand{\AlgoOutput}[1]{\AlgoMeta{Output}{#1}}
\newcommand{\AlgoRet}[1]{\AlgoLine{\textbf{return} #1}}
\newcommand{\AlgoIf}[2]{%
  \AlgoLine{\textbf{if} #1 \textbf{then}}%
  \addtolength{\AlgoIndent}{1.25em}#2%
  \addtolength{\AlgoIndent}{-1.25em}%
  \AlgoLine{\textbf{end if}}%
}
\newcommand{\AlgoEIf}[3]{%
  \AlgoLine{\textbf{if} #1 \textbf{then}}%
  \addtolength{\AlgoIndent}{1.25em}#2%
  \addtolength{\AlgoIndent}{-1.25em}%
  \AlgoLine{\textbf{else}}%
  \addtolength{\AlgoIndent}{1.25em}#3%
  \addtolength{\AlgoIndent}{-1.25em}%
  \AlgoLine{\textbf{end if}}%
}
\newcommand{\AlgoWhile}[2]{%
  \AlgoLine{\textbf{while} #1 \textbf{do}}%
  \addtolength{\AlgoIndent}{1.25em}#2%
  \addtolength{\AlgoIndent}{-1.25em}%
  \AlgoLine{\textbf{end while}}%
}

\makeatother

\title{Kozuchi Agent: A Language-Agnostic Open-Weight Agent for Software Repair}

\author{Mehdi Bahrami}
\affiliation{%
  \institution{Fujitsu Research of America}
  \city{Santa Clara}
  \state{CA}
  \country{USA}
}

\author{Kosaku Kimura}
\affiliation{%
  \institution{Fujitsu Research}
  \city{Kawasaki}
  \country{Japan}
}

\author{Satoshi Munakata}
\affiliation{%
  \institution{Fujitsu Research}
  \city{Kawasaki}
  \country{Japan}
}

\author{Satoshi Nakashima}
\affiliation{%
  \institution{Fujitsu Research}
  \city{Kawasaki}
  \country{Japan}
}

\author{Yu Ishikawa}
\affiliation{%
  \institution{Fujitsu Research}
  \city{Kawasaki}
  \country{Japan}
}

\author{Kosuke Maeda}
\affiliation{%
  \institution{Fujitsu Research}
  \city{Kawasaki}
  \country{Japan}
}

\author{Nao Soma}
\affiliation{%
  \institution{Fujitsu Research}
  \city{Kawasaki}
  \country{Japan}
}

\author{Kenichi Kobayashi}
\affiliation{%
  \institution{Fujitsu Research}
  \city{Kawasaki}
  \country{Japan}
}

\author{Keisuke Miyazaki}
\affiliation{%
  \institution{Fujitsu Research}
  \city{Kawasaki}
  \country{Japan}
}

\author{Keizo Kato}
\affiliation{%
  \institution{Fujitsu Research}
  \city{Kawasaki}
  \country{Japan}
}

\author{Shigeki Fukuta}
\affiliation{%
  \institution{Fujitsu Research}
  \city{Kawasaki}
  \country{Japan}
}

\author{Tatsuo Kumano}
\affiliation{%
  \institution{Fujitsu Research}
  \city{Kawasaki}
  \country{Japan}
}

\author{Nobutaka Imamura}
\affiliation{%
  \institution{Fujitsu Research}
  \city{Kawasaki}
  \country{Japan}
}

\author{Kevin Musgrave}
\affiliation{%
  \institution{Fujitsu Research of America}
  \city{Santa Clara}
  \state{CA}
  \country{USA}
}

\author{Shahbaz Abdul Khader}
\affiliation{%
  \institution{Fujitsu Research of Europe}
  \city{Slough}
  \country{UK}
}

\author{Kwun Ho Ngan}
\affiliation{%
  \institution{Fujitsu Research of Europe}
  \city{Slough}
  \country{UK}
}

\author{Joe Townsend}
\affiliation{%
  \institution{Fujitsu Research of Europe}
  \city{Slough}
  \country{UK}
}

\author{Fayas Asharindavida}
\affiliation{%
  \institution{Fujitsu Research of Europe}
  \city{Slough}
  \country{UK}
}

\author{Matthieu Parizy}
\affiliation{%
  \institution{Fujitsu Research of Europe}
  \city{Slough}
  \country{UK}
}

\author{Akira Sakai}
\affiliation{%
  \institution{Fujitsu Research}
  \city{Kawasaki}
  \country{Japan}
}

\author{Yuma Ichikawa}
\affiliation{%
  \institution{Fujitsu Research}
  \city{Kawasaki}
  \country{Japan}
}

\author{Yang Zhao}
\affiliation{%
  \institution{Fujitsu Research \& Development}
  \city{Beijing}
  \country{China}
}

\author{Michiaki Takizawa}
\affiliation{%
  \institution{Fujitsu Research}
  \city{Kawasaki}
  \country{Japan}
}

\author{Taku Fukui}
\affiliation{%
  \institution{Fujitsu Research}
  \city{Kawasaki}
  \country{Japan}
}

\author{Hiroki Ohtsuji}
\affiliation{%
  \institution{Fujitsu Research}
  \city{Kawasaki}
  \country{Japan}
}

\author{Wei-Peng Chen}
\affiliation{%
  \institution{Fujitsu Research of America}
  \city{Santa Clara}
  \state{CA}
  \country{USA}
}

\author{Hiromichi Kobashi}
\affiliation{%
  \institution{Fujitsu Research}
  \city{Kawasaki}
  \country{Japan}
}

\renewcommand{\shortauthors}{Bahrami et al.}



\begin{document}

\begin{abstract}
Industrial software-engineering teams increasingly need LLM agents that
turn bug reports into correct patches, yet benchmark-scale operation adds
long horizons, tool-use discipline, context persistence, heterogeneous
clusters, and evaluation reuse. We present Kozuchi Agent, a
\textit{language-agnostic open-weight repair agent} and \textit{CI-operated
evaluation pipeline}. Explicit phases, persistent state, deterministic
tools, a model-independent action interface, and cross-agent test-time
selection make runs auditable and repeatable. With locally hosted
Qwen3.5-27B, no fine-tuning, and TTS@8, Kozuchi resolves 374/500
SWE-bench Verified instances on the official evaluator. Unchanged on
Multi-SWE-bench Java, the same 27-billion-parameter agent resolves 41/128
instances (\SI{32.03}{\percent}), ranking first among strict open-weight
submissions and fourth of 42 overall; on Python it ranks 12th of 135 and
first among open-weight systems. Per-phase behavior remains within
$\pm 5$ percentage points across languages. Remaining failures mainly
reflect semantic correctness, Java-specific harness issues, and selection
errors. Across both tracks, results compare favorably with open/local
peers by parameter count. Analysis of candidate diversity, selector
regret, and patch reliability shows that the remaining gap is primarily
semantic correctness and selection rather than edit formatting or
proprietary-model access.
Operationally, reusable CI stages reduce operator touch-points from five
to one across heterogeneous internal clusters.
\end{abstract}

\maketitle

\section{Introduction}
\label{sec:intro}
\begin{figure}[t]
  \centering
  \includegraphics[width=\columnwidth]{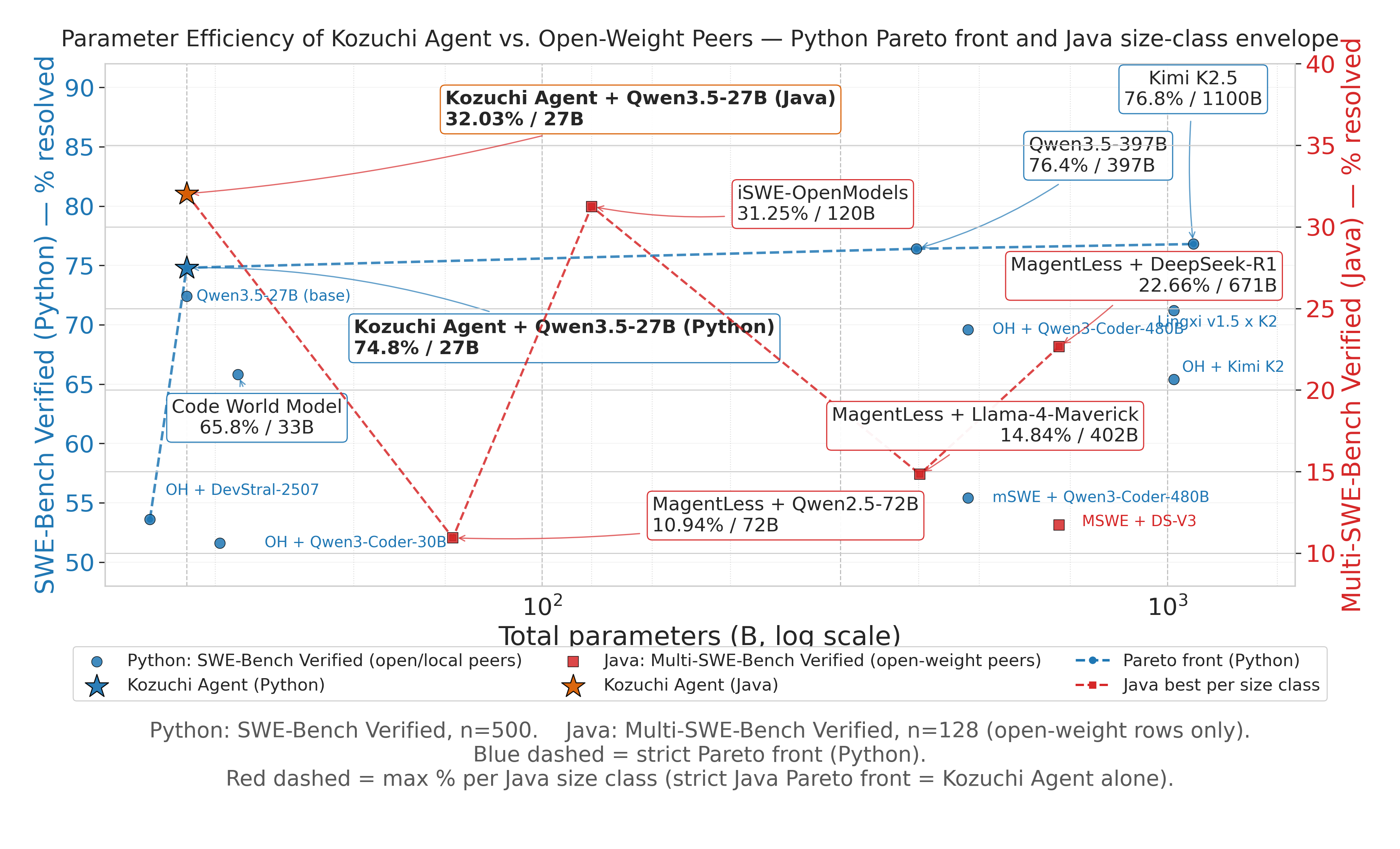}
  \caption{Kozuchi Agent is parameter efficient on both Python
  SWE-bench Verified and Multi-SWE-bench Java.}
  \Description{Dual-axis scatter plot of agent success rate against
  total parameter count on a log scale. The left (blue) axis shows
  SWE-bench Verified Python results for Kozuchi Agent and open/local
  peers with the Python Pareto frontier as a dashed blue line; the
  right (red) axis shows Multi-SWE-Bench Verified Java results for
  open-weight peers with the per-size-class Java envelope as a dashed
  red line. Kozuchi Agent + Qwen3.5-27B is highlighted with star
  markers on both axes (74.8 percent Python, 32.03 percent Java) at
  27 billion parameters.}
  \label{fig:pareto}
\end{figure}

Across our engineering organization, we observe a recurring
industrial demand: developers prefer an LLM-based agent that takes an
issue description plus a repository state and emits a correct,
test-passing patch. In SWE-bench-style
evaluations~\cite{jimenez2024swebench}, meeting that demand requires
more than choosing a model~\cite{Zhengetal2024}: the surrounding harness must keep long,
tool-using repair attempts auditable and repeatable. The central
question for us became:
\emph{how far can an open, locally hosted model go when embedded in an
auditable repair harness and selected through cross-agent testing?}
Figure~\ref{fig:pareto} summarizes the resulting Python and Java
position against open/local peers by parameter count.
Four operational blockers shaped the design as follows.

\paragraph{Pain~1: Long-Horizon Execution.}
A real fix requires reproducing the bug, generating a
regression test, locating the root cause, editing several files, and
verifying the result~\cite{jimenez2024swebench,yang2024sweagent,yao2023react,shinn2023reflexion},
with each step taking many tens of LLM turns.
In practice software-repair benchmark trajectories consume most of a multi-hour
budget. Free-form planning at that horizon degrades decision quality:
agents skip reproduction, re-investigate locations they
analyzed, or ``fix'' code they have not understood.

\paragraph{Pain~2: Tool-Grammar Drift Across Model Families.}
LLM families vary in conventions for tool calls and structured actions.
A small syntax mismatch can invalidate a reasonable
action~\cite{openaifunctioncalling2026,anthropictooluse2026} and cascade
into wasted turns. The harness must absorb
that diversity rather than forcing every model integration to fork the
agent.

\paragraph{Pain~3: Multi-Cluster Heterogeneity.}
Our jobs land on heterogeneous internal clusters: GPU inference and
training, VM-based testing, and Docker-based benchmark grading all have
different operational requirements. We need one runtime to work across
these environments without forcing researchers to rewrite scripts for
each hardware backend.

\paragraph{Pain~4: Evaluation Cost.}
Re-running SWE-bench Verified end-to-end with Docker-based grading is
expensive enough that a CI pipeline that re-runs every stage is
unusable in practice~\cite{openai2024swebenchverified,sbcli2026}. The
evaluator-grade score is the gate for
external claims and the floor for experimentation, but it
must be reachable on a budget that fits a few SLURM
jobs~\cite{yoo2003slurm} rather than a research-cluster reservation.
The CI reuse mode declared in \code{.gitlab-ci.yml} short-circuits 6 of
9 stages per iteration (\S\ref{sec:lessons}).

This paper is an experience report on the harness and pipeline we built
to address those four pains, and on what we learned operating it. Its
industrial contribution is the workflow replacement: multi-cluster
orchestration, reusable artifacts, trajectory-level audit, and one-push
benchmark reruns.
\emph{Kozuchi Agent} comprises:
\begin{itemize}
\item a phase-driven SWE agent layered on \code{mini-swe-agent}, a
minimal open-source agent runtime~\cite{miniswea};
\item a fixed sandbox of dynamic-analysis and edit operations;
\item a CI-driven multi-cluster pipeline for repeatable pipelines;
\item a test-time selection layer that merges multiple inference
runs through \emph{cross-agent testing}~\cite{komoravolu2026agent}.
\end{itemize}
The released artifacts expose the resulting trajectories, selector
choices, evaluation outputs, and additional deep analysis.

\textbf{Scope of industrial claims.}
The reported industrial impact is the internal evaluation-pipeline
replacement described in Table~\ref{tab:workflow-replacement} (touch-points,
reusable stages, cluster abstraction); we do not report production
traffic, end-user developer KPIs, or deployment of the agent against
proprietary internal repositories. The strongest causal evidence in this
submission isolates candidate count and cross-agent selection; phase,
handover, formatter, tool, and CI-reuse mechanisms are supported by
operational signatures and artifact-level audits rather than
component-removal reruns (Table~\ref{tab:evidence-map}).

\textbf{Contributions.}
\noindent\textbf{C1. A phase-driven, model-agnostic SWE agent harness with auditable tools.}
We decompose issue resolution into eight semantically meaningful phases
connected by explicit success and fallback transitions, separate the
logical action contract from model-specific surface syntax, and expose a
small deterministic tool sandbox for dynamic tracing, caller discovery,
and guarded editing. These design mechanisms make a free-form repair
loop inspectable: state can be resumed, audited, and compared across
open-weight and API-served backends. C1 is supported by operational
signatures (\S\ref{sec:observations}, \S\ref{sec:design}) and
artifact-level audit (Data Availability Statement), not by mechanism-removal
ablations; controlled removals of the phase graph, formatter,
persistent state, and tool sandbox are scoped as future work
(\S\ref{sec:conclusion}). Table~\ref{tab:evidence-map} summarizes this
evidence boundary for all mechanisms and claims.

\noindent\textbf{C2. A CI-operated multi-cluster benchmark pipeline with reusable and regenerable artifacts.}
A single CI pipeline coordinates benchmark inference, grading, data
synthesis, optional post-training, test-time selection, and reporting
across heterogeneous clusters. Its reuse mode records which expensive
stages can be short-circuited using previously validated outputs, while
the published analysis scripts in \code{src/} regenerate every
empirical value cited in RQ1--RQ6 directly from the SWE-bench and
Multi-SWE-bench Java harness reports, trajectories, and cross-agent
selection bundles in the public artifact (Data Availability Statement).

\noindent\textbf{C3. A cross-agent test-time selector with trajectory-level audit.}
We merge $K=8$ candidate inference runs through cross-agent testing:
each run contributes both a patch and independent tests, and the tests
are cross-applied to score candidate patches without using the hidden
benchmark tests or a learned verifier. We also analyze the eight
candidate streams and 495 full message trajectories, separating selector
regret from
instance hardness (408 instances solved by at least one run, 234 by all
runs, 92 by none) and surfacing content-level selector signals such as
\code{THOUGHT:}-block-to-action balance, rollback language, and
error-marker asymmetries.

\noindent\textbf{C4. Competitive open-weight repair results with statistical and failure-mode analysis.}
On the official SWE-bench Verified cloud evaluator, our 27-billion-parameter open-weight
Python configuration resolves 374 of 500 instances (\SI{74.80}{\percent}, Wilson \SI{95}{\percent} CI [\SI{70.82}{\percent},
\SI{78.41}{\percent}]). The internal Docker re-grade resolves 376 (\SI{75.2}{\percent}); the
cross-agent selector contributes +14 resolved instances over a
size-ordered first-candidate baseline. On a frozen snapshot of 135
publicly catalogued SWE-bench
Verified submissions the configuration ranks 12th overall and is the
highest-ranked open-weight system; under Benjamini--Hochberg FDR
correction (target $q{\le}0.05$) the configuration significantly
outperforms 16 of 17 curated open-weight peers on matched
per-instance McNemar tests. Of 126 unresolved instances, \SI{91.3}{\percent} are
patches that apply cleanly but
do not flip the hidden tests; \SI{0}{\percent} are malformed-diff or empty-patch
failures, isolating the residual gap as a semantic-correctness problem
rather than an editing-format problem. Patch LOC churn is the strongest
single feature signal (point-biserial $r{=}-0.197$,
$p{=}1.1{\times}10^{-5}$), and 494/495 produced patches apply cleanly
through the SWE-bench harness. On Multi-SWE-bench Java, the
unchanged 27-billion-parameter Java configuration resolves 41/128 instances (\SI{32.03}{\percent}), ranks
4 of 42 overall, and is first among strict open-weight submissions;
under Benjamini--Hochberg FDR control ($q\le0.05$), it significantly
outperforms 34 of 41 Java peer systems on paired per-instance outcomes.

\noindent\textbf{C5. Cross-language transfer of the same harness to Java.}
The phase decomposition, action contract, persistent shared state, and
cross-agent selector transfer from Python to Java unchanged. Per-phase
share of assistant messages matches the Python run within $\pm5$
percentage points on every phase, six of seven per-instance effort
medians sit within \SI{20}{\percent} of Python parity, and the same 27-billion-parameter Qwen3.5
backbone beats the strongest same-class open-weight Java peer
(Qwen2.5-72B) by 38 instances, compared with a 26-instance same-family
margin on Python.

\noindent\textbf{Evidence boundary.}
Table~\ref{tab:evidence-map} separates what this experience report
establishes through controlled comparisons from what it supports only
through operational signatures. ``Ablated'' rows have a matched
comparison in the published artifact; ``operational'' rows are
evidenced by trajectory audits and inventories without
component-removal reruns; ``estimate'' and ``out of scope'' rows are
disclosed as such wherever they are cited.

\textbf{Roadmap.}
\S\ref{sec:related}--\S\ref{sec:definitions} cover prior systems,
industrial background, and the runtime definitions. \S\ref{sec:observations}--\S\ref{sec:design-tts}
describe the observations, harness design, and cross-agent selector.
\S\ref{sec:eval} reports six scripted analyses over the published
Python TTS@8 and Java xcheck@8 artifacts. \S\ref{sec:lessons} states
operational lessons, \S\ref{sec:conclusion} concludes, and
the Data Availability Statement defines the public artifact boundary.

\section{Related Work}
\label{sec:related}

\paragraph{Agentic SWE Systems.}
SWE-agent~\cite{yang2024sweagent} established the now-standard pattern
of an LLM observing a shell and emitting commands, building on broader
reason-act and reflective-agent patterns~\cite{yao2023react,shinn2023reflexion,madaan2023selfrefine,yao2023tree}.
We build on a minimal open-source runtime and add a phase-driven contract,
a fixed tool sandbox, and a multi-cluster pipeline. AGENTLESS~\cite{xia2024agentless}
uses a simpler localization-fix-verify pipeline; OpenHands and
AutoCodeRover provide complementary open software-agent platforms and
program-improvement systems~\cite{wang2024openhands,zhang2024autocoderover}.
\begin{table}[t]
\centering
\caption{Evidence map: mechanisms and claims by type.}
\label{tab:evidence-map}
\begin{tabular}{@{}p{0.4\columnwidth}p{0.36\columnwidth}l@{}}
\toprule
Mechanism / claim & Evidence type & Where \\
\midrule
Candidate count $K{=}8$ & Ablated: pass@1 334--343 vs.\ oracle 408 & Tab.~\ref{tab:selector-ablation} \\
Cross-agent selector & Ablated: 376 vs.\ 362 (order) vs.\ 346 (self-tests) & Tab.~\ref{tab:selector-ablation} \\
Selector weights $w_B,w_R$ & Sensitivity sweep: $\le$5/500 outcome change & \S\ref{sec:design-tts} \\
Phase graph, handover state, formatter, tool sandbox, CI reuse & Operational signatures; not ablated & \S\ref{sec:design}, \S\ref{sec:lessons} \\
Engineer-minute saving & Derived order-of-magnitude estimate & Tab.~\ref{tab:workflow-replacement} \\
GPU-hours & Wall-clock-derived estimate & \S\ref{subsec:rq5} \\
Production deployment, developer pilot, GPU accounting & Out of scope; future work & \S\ref{sec:conclusion} \\
\bottomrule
\end{tabular}
\end{table}
Our system is closer to a long agentic loop but constrains it through
phases and shared state.
The clearest difference is how test assets are handled: AGENTLESS selects
one reproduction test and one regression set per issue, whereas each
Kozuchi agent run independently generates tests that are later cross-applied to
other candidates.

\paragraph{Industry-Facing Coding Agents.}
Commercial systems such as GitHub Copilot's cloud coding
agent~\cite{githubcopilotagent2025}, Amazon Q Developer~\cite{amazonqdeveloper2024},
Google Jules~\cite{googlejules2025}, OpenAI Codex~\cite{openaicodex2025},
Anthropic Claude Code~\cite{anthropicclaudecode2025}, and
Devin~\cite{devin2024} validate industrial demand for asynchronous
issue-to-patch agents. Their public descriptions are necessarily
product-oriented; our focus is complementary harness evidence: phase
boundaries, deterministic tools, queue-aware CI orchestration, reusable
benchmark artifacts, and inspectable selection.

\paragraph{Tools, Training, and CI.}
Toolformer-style work injects tool calls into model training~\cite{schick2023toolformer}; generated-test
selection and pass@k-style code evaluation provide related precedents for
using execution evidence around generated code~\cite{chen2021codex,chen2023codet}.
Our approach is the dual, holding the model fixed while injecting classic SE
tools (line tracing, caller discovery, guarded editing) at runtime. SFT
and RL recipes for code agents remain active directions, and systems such
as Nemotron-CORTEXA~\cite{cortexa2025} use fine-tuned retrievers and
graph localization. R2E-Gym~\cite{r2egym2025} explores verifiers for
BEST@K; our cross-agent testing is execution-based but deliberately
model-free. Finally, deployed LLM-for-CI systems such as
LogSage~\cite{logsage2025} share our concern for artifact availability
and reproducibility, although in our case CI orchestrates the agent
workload itself.

In sum, the individual harness mechanisms---phased workflows, persistent
state, runtime tools, CI orchestration---draw on established practice.
The originality of this work lies in the cross-agent selector
formulation, in which independent runs generate tests that are
cross-applied as selection evidence under a constraint that no hidden or
production-only feedback is available at selection time, and in the
integration and industrial-scale evaluation of these mechanisms as one
auditable harness.

\section{Background and Industrial Context}
\label{sec:background}

\subsection{SWE-bench and Multi-SWE-bench Java}
SWE-bench is a benchmark of real GitHub issues paired with the
affected repository state and a hidden test suite; an entry counts as
\emph{resolved} only when the model's patch passes the issue-revealing
tests while preserving the existing regression tests~\cite{jimenez2024swebench}.
It extends a longer code-generation and program-repair evaluation line
from function-level benchmarks to repository-level issues~\cite{chen2021codex,hendrycks2021apps,li2022alphacode,legoues2012genprog,long2016prophet,gazzola2019automatic}.
SWE-bench Verified is a 500-instance human-curated subset with cleaner
problem statements and reliable evaluation
tests~\cite{openai2024swebenchverified}; it is the Python benchmark used
in RQ1--RQ5.  Multi-SWE-bench Java is the corresponding Java
track used in RQ6: it keeps the same issue-to-patch resolved metric, but
evaluates Java repositories through Maven/Gradle-style harnesses and a
separate public leaderboard~\cite{zan2025multiswebench}.  We report the Python result on the
\code{sb-cli} cloud stack~\cite{sbcli2026} and the Java result on the
Multi-SWE-bench Java leaderboard, keeping each benchmark's
denominator and peer set separate.

\subsection{Agentic Loops and Base Runtime}
Kozuchi Agent uses \code{mini-swe-agent}~\cite{miniswea} as its base
tool-using repair loop. Its contribution is the harness around that
loop: phase control, persistent shared state, deterministic tools, and
model-independent action formatting for long repairs.

\subsection{The Workflow We Set Out to Replace}
Before Kozuchi Agent, each iteration crossed heterogeneous internal
clusters and required manually starting a vLLM server, launching a
one-off repair-benchmark run, waiting hours, running Docker grading, and
recording scores in a spreadsheet. The harness replaces this workflow
with a CI-driven pipeline, one configurable agent, a fixed runtime
contract, and a publishable artifact bundle. Table~\ref{tab:workflow-replacement}
summarizes this workflow inventory; the engineer-minute estimate is an
order-of-magnitude operational estimate bounded by published SWE-bench
Docker-runtime and CI-adoption studies~\cite{epoch2024swebenchdocker,hilton2016ci}.
\begin{table}[t]
\centering
\caption{Per-evaluation-cycle workflow comparison.}
\label{tab:workflow-replacement}
\begin{tabular}{@{}p{0.34\columnwidth}p{0.27\columnwidth}p{0.31\columnwidth}@{}}
\toprule
Aspect & Pre-Kozuchi agent & Kozuchi agent \\
\midrule
Operator touch-points / cycle & 5 (vLLM, launch, wait, grade, log) & 1 (CI push) \\
Operator effort / cycle (est. min) & \(\approx\)75 & \(\approx\)5 \\
Backend integrations & separate integration effort & 4 action formats \(\times\) 17 model configs \\
Reusable pipeline stages & 0 of 9 & 6 of 9 \\
Cluster targets / abstraction & per-cluster scripts & 5 (\code{ENV\_NAME}) \\
Score provenance & manual score log & 495 traj.\ + selection JSON \\
Patch applicability & not measured & 494/495 clean apply \\
\bottomrule
\end{tabular}

\end{table}


\section{Definitions and Design Invariants}
\label{sec:definitions}

We adopt the following definitions for the rest of the paper. Some
invariants are mechanically enforced by parsing and configured
transitions; others are prompt/workflow contracts audited through
trajectories.

\paragraph{D1 (Agent Runtime $\mathcal{R}$).}
The runtime is
$\mathcal{R}=\langle\mathcal{P},\mathcal{S},\mathcal{T},\mathcal{A},\mathcal{F},\mathcal{H}\rangle$:
phases, shared state, tools, action format, parser, and handover
function. This tuple is declared by configuration rather than recovered
from ad hoc scripts.

\paragraph{D2 (Trajectory).}
A trajectory is the ordered record of prompts, assistant completions,
parsed tool calls, tool responses, format/timeout feedback, phase
handovers, and shared-state snapshots. Trajectories are published per
instance and are the main audit object.

\paragraph{D3 (Phase Transition System).}
The phase transition system is a directed graph $G_\Phi=(V_\Phi,E_\Phi)$.
The evaluated graph has eight phases and fifteen normal/fallback edges
(\autoref{alg:phase-transition}).

\paragraph{D4 (Tool-Call Contract).}
Each assistant turn must contain exactly one parseable action. Four
surface formats let the same logical contract serve different model
families while keeping parser failures explicit in the trajectory.

\paragraph{D5 (Verifier/Selector).}
A verifier returns a binary or numeric pass-rate signal for a candidate
patch. We use an intra-trajectory verifier for tests generated inside one
run, and a cross-agent selector (\S\ref{sec:design-tts}) that
cross-applies each run's generated tests to every candidate patch.

\paragraph{D6 (Artifact Boundary).}
The public artifact includes trajectories, patches, per-run reports,
selection tables, and official submission files. Transient scratch state,
internal cluster configuration, and pipeline source are outside the
release boundary because they contain internal paths, runners, tokens, or
licence-sensitive dependencies.

\paragraph{Runtime Invariants.}
Every assistant turn yields one parseable action or a recorded format
error; phase exits follow configured edges; generated notes and logs stay
outside the submitted patch; long observations are summarized with
explicit elision; and repeated failures or repeated identical actions
trigger a strategy-change warning.

\begin{algorithm}[t]
  \caption{Phase-driven repair orchestration}
  \label{alg:phase-transition}
  \AlgoInput{Issue $I$, repository checkout $R$, phase graph $\Phi$, skill library $\mathcal{K}$, tool set $\mathcal{T}$, action grammar $A$, shared artifact store $S$, and limits $B$.}
  \AlgoOutput{Candidate patch $\Delta$, final report $\rho$, and trajectory bundle $\tau$.}
  \AlgoStmt{$p\gets p_{\mathsf{reproduce}}$; $H\gets\emptyset$; $M\gets\emptyset$; $\tau\gets\emptyset$}
  \AlgoNote{In the evaluated graph, $p$ ranges over issue reproduction, FAIL-to-PASS test synthesis, code localization, PASS-to-PASS test localization, code fixing, patch verification, issue closeout, and final reporting.}
  \AlgoWhile{$p\neq p_{\mathsf{final}}$}{
    \AlgoStmt{$K_p\gets\{k\in\mathcal{K}: p\in\mathsf{phases}(k)\land\mathsf{tools}(k)\subseteq\mathcal{T}\}$}
    \AlgoStmt{$C_p\gets\mathsf{Render}(I,R,\Phi[p],K_p,S,M,H)$}
    \AlgoNote{$C_p$ contains the phase definition, available tools, selected skill guidance, handover memos, and artifacts already written to $S$.}
    \AlgoStmt{$e\gets\bot$}
    \AlgoWhile{$e=\bot$}{
      \AlgoStmt{$y\gets\mathsf{LLM}(C_p,H)$}
      \AlgoStmt{$a\gets\mathsf{Parse}_{A}(y)$}
      \AlgoIf{$a=\bot$}{
        \AlgoStmt{$o\gets\mathsf{FormatError}(y,A)$}
        \AlgoNote{Ambiguous or multi-action output is recorded but not executed.}
      }
      \AlgoIf{$a\neq\bot$}{
        \AlgoStmt{$o\gets\mathsf{SandboxExec}(a,R,\mathcal{T})$}
        \AlgoStmt{$S\gets\mathsf{UpdateArtifacts}(S,p,a,o)$}
        \AlgoNote{Phase artifacts include reproduction scripts, tests, traces, root-cause notes, patch diffs, and reports.}
      }
      \AlgoStmt{$H\gets H\circ(y,a,o)$; $\tau\gets\tau\circ(p,y,a,o,S)$}
      \AlgoStmt{$e\gets\mathsf{ExitMarker}(y,o)$}
      \AlgoIf{$e=\bot$ and $\mathsf{OverBudget}(H,S,B,p)$}{
        \AlgoStmt{$m\gets\mathsf{Compress}(H,S,p)$; $S\gets S\cup\{m\}$; $H\gets\emptyset$}
        \AlgoStmt{$C_p\gets\mathsf{Render}(I,R,\Phi[p],K_p,S,M,H)$}
      }
    }
    \AlgoStmt{$(p,M)\gets\mathsf{PhaseHandover}(\Phi,p,e,H,S,M)$}
    \AlgoNote{Handover extracts phase artifacts, follows the configured next/fallback edge, writes a memo, and preserves scratch state outside the patch.}
  }
  \AlgoStmt{$(\Delta,\rho)\gets\mathsf{CollectFinalArtifacts}(S,M)$}
  \AlgoRet{$(\Delta,\rho,\tau)$}
\end{algorithm}


\section{Motivating Empirical Observations}
\label{sec:observations}

Early prototypes exposed recurring operational failures: long repair
contexts lost evidence and repeated work, model-specific tool syntax made
backend swaps brittle, shell-only debugging localized defects poorly, and
multi-hour benchmark stages made recomputation expensive.  The system
design below turns these observations into phase boundaries, durable
shared state, deterministic tools, configurable action formats, and CI
artifact reuse.

\section{System Design}
\label{sec:design}

\begin{figure*}[t]
  \centering
  \includegraphics[width=0.72\textwidth]{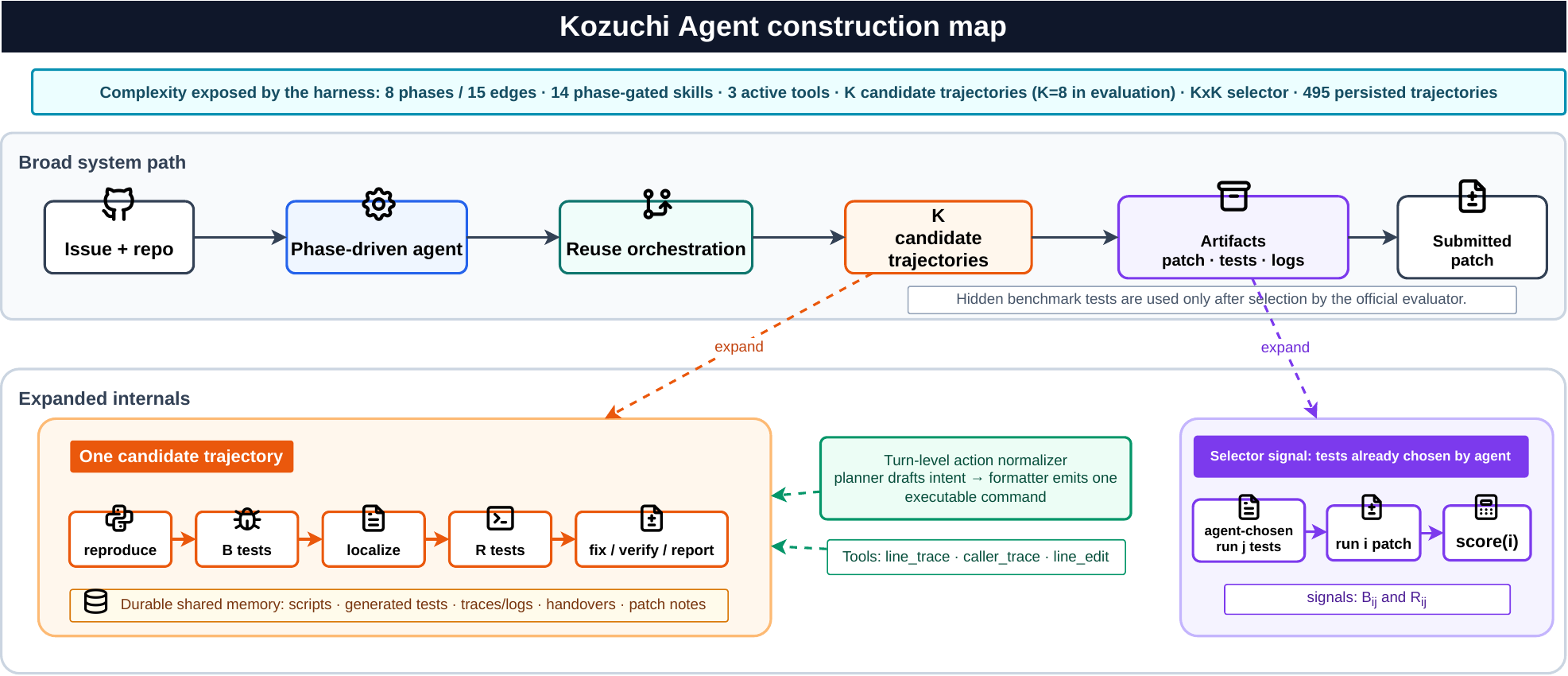}
  \caption{Kozuchi Agent workflow from candidate generation to
  cross-agent selection.}
  \Description{Two-lane system diagram. The upper lane shows the broad
  issue-to-patch workflow through candidate generation, archived
  artifacts, selection, and submitted patch. The lower lane expands one
  candidate trajectory with durable shared memory, turn-level action
  normalization, tools, and the cross-run test application that produces
  the selector signals B_ij and R_ij.}
  \label{fig:arch}
\end{figure*}

Figure~\ref{fig:arch} summarizes the architecture at two connected
resolutions. The broad path shows how the pipeline
turns one issue into $K$ auditable candidate trajectories and a final
submitted patch. The in-depth path shows the internal structure that
makes those trajectories useful evidence: phase boundaries define what
each run must produce, the agent chooses or synthesizes the tests during
those phases, durable shared memory persists tests and handover state,
deterministic tools make localization and edits inspectable, and
cross-agent selection converts those independently chosen tests into signals as defined in Eq~\S\ref{eq:score}.

\subsection{Agent Runtime Contract}
\label{sec:design-runtime}

The agent runtime is specified declaratively: one configuration file
defines the phase graph, prompt templates, action-interface family,
prompt-guidance skill library, and operational limits. The runtime
contract is therefore explicit before execution, instead of being
recoverable only from scattered scripts.

\paragraph{Phases.}
The evaluated phase graph has eight nodes and fifteen directed edges
(\autoref{alg:phase-transition}). Each phase encodes a local workflow and two
possible exits: successful completion or explicit give-up. This
structure does not make the agent more capable by itself; rather, it is
designed to keep long trajectories from becoming unstructured
conversations.
The phase definitions and handover prompts make phase exits auditable:
when a run leaves a phase, the next phase is selected from the
configured edge and the handover memo records what evidence is carried
forward.  Concretely, \textsc{PhaseHandover} extracts scripts, traces,
tests, patch diffs, verification logs, and reports from the exiting
phase; chooses either the configured next edge or the configured fallback
edge based on the exit reason; writes a memo that summarizes carried
evidence and open risks; and preserves scratch notes in shared state
rather than in the submitted patch.

\paragraph{Proposition 1 (Workflow Well-Formedness).}
\textit{In the observed Python headline trajectories, phase visits use the
configured phase names and the run terminates either in final reporting
or in an explicit runtime outcome.}

\emph{Argument.} The phase graph is declared in configuration, and the
trajectory-derived phase summaries show complete coverage of the eight
declared phases for the 495 persisted trajectories. We do not claim
liveness---the agent is not guaranteed to solve every instance---only
that the published run is auditable against the configured phase graph.

\paragraph{Skills (Phase-Gated, Tool-Gated).}
Skills are reusable prompt-guidance blocks, not learned model
capabilities. Each skill gives the agent a small procedure or tool-use
pattern, and is injected only when the current phase and available tools
make it relevant. The current configuration contains 14 such
phase- and tool-gated skills. This deliberately conservative injection
policy shortens the prompt and narrows the action space, which matters
when a repair trajectory can span hundreds of turns.

\paragraph{Action Formats.}
The action interface separates the semantic requirement---one executable
action per turn---from model-specific syntax. The same runtime can
therefore evaluate several model families without changing the agent's
state machine or tool semantics.

\paragraph{Tool Sandbox.}
At runtime, the agent receives a fixed set of deterministic
software-engineering operations. The current set covers dynamic line
tracing, caller discovery, and guarded line editing. Keeping this set
small made tool behavior inspectable and prevented the agent from
depending on a large, brittle tool zoo.

\paragraph{Templates.}
Templates enforce the interaction contract: each turn contains an
observable rationale field and exactly one action, and long observations
are summarized with explicit elision rather than silently dropped. We
treat that field as trajectory text, not as a faithful record of the
model's cognitive state; the goal is to make the observable interface
stable.

\paragraph{Context Compression and Handover.}
State handover addresses a simple failure mode: facts stored only in the
conversation can disappear when context is compressed. At phase changes,
and when a phase grows too long, the agent writes a concise memo and
continues with the relevant shared artifacts. The filesystem, not the
raw conversation, becomes the durable memory.

\paragraph{Orchestra Runtime.}
For the Python headline submission, each turn is processed by two roles. A
planning role drafts the next step, and a formatting role rewrites only
the executable action to satisfy the active interface. Both roles use
the same locally hosted model. This division is a
harness-level substitute for fine-tuning: the planner can express the
repair strategy naturally, while the formatter normalizes only the
command surface before execution, reducing formatting failures without
changing model weights.

\subsection{Software-Engineering Tool Suite}
\label{sec:design-tools}

The suite is deliberately small: \textbf{line tracing} records which
statements and values appear under a reproduction test, turning
localization into execution evidence; \textbf{caller tracing} shows the
paths that reach suspect code and connects local symptoms to
user-visible behavior; and \textbf{guarded editing} couples each edit to
an expected target line, avoiding many free-form patch corruptions. The
main policy is phase-gated injection: the agent sees only the tools and
guidance relevant to its current phase, making tool availability part of
the experimental contract.

\subsection{Multi-Cluster Pipeline}
\label{sec:design-pipeline}

The pipeline turns benchmark experimentation into a set of reusable
stages: candidate generation, grading, data synthesis, optional
post-training, test-time selection, and reporting. Each stage can run
locally for debugging or as a scheduled cluster job. The practical
contribution is reuse: operators can re-run only the stage under study
while holding earlier artifacts fixed. This is what made repeated
TTS@8 sweeps feasible in day-to-day development.

\subsection{Model-Agnostic Serving}
\label{sec:design-serving}

The serving layer decouples model identity, chat formatting, and
execution sandbox. The same agent can run against local
open-weight models, multi-server vLLM deployments, and remote APIs~\cite{kwon2023vllm}. The
backend inventory is reported with the artifact materials.

\subsection{CI as the Evaluation Substrate}
\label{sec:design-ci}

CI is the evaluation substrate, not just a test runner. It records the
stage graph, the cluster target, the reusable artifacts, and the
conditions under which expensive work should be skipped. The stage
graph covers preparation, candidate generation, grading, synthesis,
optional training, post-training benchmarking, selection, and reporting;
reuse short-circuits let operators iterate on one stage without
invalidating the rest. This follows standard CI practice of keeping
frequent integration feedback cheap and repeatable~\cite{fowler2006ci,duvall2007ci}.

\section{Cross-Agent Test-Time Selection}
\label{sec:design-tts}

A central design choice is to separate \emph{candidate generation}
from \emph{final selection}. With $K=8$ independent runs of the same
locally hosted model, each run produces a candidate patch
together with its own bug-revealing and regression-preservation tests.
The candidate agent decides which tests exist before final selection;
the cross-agent selector does not choose tests, but only cross-applies the
archived test runners. Each run contributes two kinds of evidence:
tests intended to expose the reported bug, and tests intended to guard
nearby behavior against regressions. These tests are fixed before final
selection, so the selector can compare candidate patches without
accessing hidden benchmark tests.
At final selection time, we form a $K\times K$ matrix in which entry
$(i,j)$ is the result of running run~$j$'s archived tests against
run~$i$'s patch, converting candidate diversity into a selection signal
using only evidence archived by the candidate runs.

\paragraph{Selector.}
For each candidate~$i$ we compute
\begin{equation}
\mathrm{score}(i)
= \frac{1}{K}\sum_{j=1}^{K}
   \big(w_{B}\, B_{ij} + w_{R}\, R_{ij}\big)
\label{eq:score}
\end{equation}
Here $B_{ij}$ and $R_{ij}$ are the pass rates obtained by the
cross-application process above: $B_{ij}$ is the pass rate of
candidate~$i$ on run~$j$'s bug-revealing
(\code{FAIL\_TO\_PASS}) tests, and $R_{ij}$ is the corresponding
pass rate on its regression (\code{PASS\_TO\_PASS}) tests. The
weights $w_{B}{=}0.3$ and $w_{R}{=}0.7$ are pragmatically selected
fixed global constants in the submitted selector artifact; they are not
learned and were not tuned on benchmark outcomes. The asymmetry encodes
a deliberate engineering prior for a merge-gated setting: a patch that
silently breaks existing behaviour (a regression that a
\code{PASS\_TO\_PASS} suite would catch) is the costlier failure once
merged, whereas agent-generated bug-revealing tests are sparser and
noisier as evidence, so regression conformance is weighted higher than
bug-revealing pass rate. Nevertheless, an artifact-only sweep over seven
nearby weight pairs changes at most 35/495 selected candidates and
5/500 estimated outcomes ($\le$\SI{1.0}{\percentagepoint}), so we report the submitted pair
as a robust operating point rather than an optimized one; a systematic
weight search on held-out tasks remains future work. We then pick
$\arg\max_i \mathrm{score}(i)$; when execution evidence is indistinguishable,
we break ties by shorter patch length as a conservative engineering heuristic. A tie
means that at least two deduplicated patches are within
$10^{-3}$ of the best score; this occurs on 304/495 covered
instances. The selector summary is reproduced as
Table~\ref{tab:xcheck}.

\begin{table}[t]
\centering
\scriptsize
\caption{Cross-agent selector summary.}
\label{tab:xcheck}
\resizebox{\columnwidth}{!}{%
\begin{tabular}{@{}lrlr@{}}
\toprule
Quantity & Value & Quantity & Value \\
\midrule
Expected (Verified)               & 500 & $w_{B}$                           & 0.3 \\
Selected (covered)                & 495 & $w_{R}$                           & 0.7 \\
Resolved (xcheck selector)        & 376 & Tie epsilon                       & 0.001 \\
Score over expected (xcheck)      & 0.752 & Tie-breaker                       & shortest\_patch\_raw \\
Resolved (order baseline)         & 362 & Tie-break applied               & 304/495 \\
Score over expected (order)       & 0.724 & Mean unique patches               & 5.59/8 \\
Tie instances                     & 304 & Matrix executions                 & 21,635 \\
Reordered by selector             & 302 &                                  & \\
\bottomrule
\end{tabular}%
}

\end{table}

\paragraph{Why Cross-Agent Rather Than Oracle Best-of-$K$.}
We do not use the official benchmark tests at selection time; doing so
would leak evaluation information. The same constraint matters in
production: a bad merge can cause significant downstream damage, so the
system needs a selection signal before any hidden or production-only
feedback is available. Instead, each run contributes the tests that its
own agent chose or generated while solving the issue.
The selector treats those archived tests as fixed evidence and
cross-applies them to other runs' patches, turning trajectory diversity
into selection evidence without any learned verifier or LLM judge.
The matrix admits only archived suites with both
\code{has\_fail\_to\_pass} and \code{has\_pass\_to\_pass}, so
empty or invalid generated-test bundles are dropped before selection
(12--21 drops per leg); candidate patches are deduplicated by content
hash, and the 2/2{,}768 deduplicated patches with non-\code{ok}
apply status are excluded from scoring. We do not infer flakiness or
over-breadth from logs; after archive creation the selector reads only
suite outcomes and apply status, never SWE-bench hidden-test results.
Per-instance
reordering data backs this up: on $302/495$ instances the selector chose
a candidate other than the order baseline, demonstrating that it
actively re-ranks rather than acting as a no-op.

\paragraph{Diversity Ceiling.}
A simple bound on selector quality comes from the union and intersection
of the eight runs: at least one run resolves 408 instances, while all
runs resolve 234. The cross-agent selector reaches 376 on the internal
Docker re-grade (374 on the official evaluator), or about
$376/408=\SI{92.2}{\percent}$ of the oracle union ceiling under the internal Docker
re-grade. Pairwise completion
overlap is high, so the useful diversity is mostly in \emph{which}
instances are solved, not in whether runs cover
different parts of SWE-bench Verified. Table~\ref{tab:tts-decomp} gives
the candidate-level decomposition that sharpens this bound: for 92
instances, none of the eight runs produces a
resolving candidate, so these instances cannot be recovered by a better
selector. Only 174 instances are ``marginal''---resolved by some but not
all runs ($1\le r_i\le 7$)---and therefore actually expose a selection
choice.
Measured against the official Python headline, the cross-agent selector leaves
34 instances (\SI{6.8}{\percentagepoint}) below the oracle union; linear interpolation on
the closed-form oracle pass@k curve places the cross-agent selector near
an oracle with $k\approx2.2$, so much of the eight-way compute budget
is currently spent on candidate generation that the selector does not
fully monetise.

\begin{table}[t]
\centering
\scriptsize
\caption{Candidate-level TTS@8 decomposition.}
\label{tab:tts-decomp}
\begin{tabular}{@{}llr@{}}
\toprule
Quantity & Interpretation & Value \\
\midrule
Mean pass@1 & Average over eight runs & \SI{67.7}{\percent} \\
All-run solved & $r_i=8$ scaffold-easy instances & 234 / 500 \\
Zero-run solved & $r_i=0$ scaffold-hard instances & 92 / 500 \\
Marginal instances & $1\le r_i\le7$ selector-sensitive & 174 / 500 \\
Oracle pass@8 & Any run resolves & 408 / 500 = \SI{81.6}{\percent} \\
Cross-agent selector & Official-evaluator result & 374 / 500 = \SI{74.8}{\percent} \\
Selector regret & Oracle minus submitted result & 34 / 500 = \SI{6.8}{\percentagepoint} \\
\bottomrule
\end{tabular}
\end{table}

\section{Evaluation}
\label{sec:eval}

\textbf{Evaluation setup.}
RQ1--RQ5 use SWE-bench Verified as the fixed Python task suite; RQ6
uses Multi-SWE-bench Java as a cross-language validation track.
The Python headline configuration uses a locally hosted model out of the box:
the approach does not require training or fine-tuning before connecting
the model to the agent runtime. For Python we run the agent with the same model eight times,
archive each run's candidate patch and tests, and then apply the
cross-agent selector described in \S\ref{sec:design-tts}. We report
the official SWE-bench cloud evaluation as the primary Python result and
use internal Docker re-grading only as a consistency check. The Java RQ6
result uses the same 27-billion-parameter agent and phase graph, but the Java xcheck@8
artifact and Java leaderboard rather than the Python Docker/cloud
evaluator. This setup separates three questions for each track: how often
the full system resolves issues, how it compares with published peers,
and where remaining errors come from.

\textbf{Reproducibility configuration.}
The public artifact records two reproducibility layers: the declarative
runtime configuration (\code{configs/agent\_sota.yaml}) and the executed
Python sweep manifest (\code{provenance/sweep\_manifest.json}). The run
uses backbone \code{Qwen/Qwen3.5-27B}~\cite{qwen35}, $K{=}8$ independent
candidate runs, selector weights $w_B{=}0.3$ and $w_R{=}0.7$
(\S\ref{sec:design-tts}), run labels \code{r01\_s1001} through
\code{r08\_s1008}, and the tool set \code{line\_trace},
\code{caller\_trace}, and \code{line\_edit}. The manifest records
serving/run parameters including two-GPU tensor parallelism, 50 shards,
\code{max\_model\_len=150000}, \code{max\_new\_tokens=16384},
\code{step\_limit=1000}, \code{instance\_timeout=06:00:00}, and the
best-known repository commit before submission; the analysis scripts use
fixed seed \code{SEED=20260427}. Model weights remain external licensed
dependencies (Data Availability Statement), so the artifact pins the model identifier and run
provenance rather than redistributing the checkpoint.

The Python analysis uses the published TTS@8 artifacts: 495 SWE-bench harness
reports, 495 trajectories, the cross-agent selection bundle, and the
per-instance outcome vectors of a curated competitor set.  We adopt the
canonical denominator $N{=}500$ for every SWE-bench Verified aggregate;
the five trajectory/report pairs that
are missing from the bundle are folded into the unresolved bucket as
\code{MISSING\_ARTEFACT} failures rather than dropped. The Java analysis
uses the published Multi-SWE-bench Java submission bundle with
canonical denominator $N{=}128$, described separately in RQ6.

\subsection{RQ1: Headline Accuracy and Uncertainty}
\label{subsec:rq1}

\textbf{Result.}
The Python configuration resolves 374/500 instances on the official cloud
evaluator (\SI{74.80}{\percent}, Wilson \SI{95}{\percent} CI [\SI{70.82}{\percent}, \SI{78.41}{\percent}]~\cite{wilson1927score}). The
corresponding internal Docker re-grade resolves 376/500 (\SI{75.2}{\percent}); the
two-instance delta reflects evaluator differences and we report the
official Python number as the Python headline. A repository-clustered bootstrap
($B{=}10{,}000$ replicates resampling at the repository level)
returns a wider \SI{95}{\percent} CI of [\SI{67.0}{\percent}, \SI{79.8}{\percent}], reflecting the fact
that resolution is correlated within repositories; we therefore
report both intervals to make the headline comparable to peers under
their own reporting conventions and conservative to clustering.

\textbf{Per-repository breakdown.}
Table~\ref{tab:byrepo} gives the repository breakdown.  The largest
repository, \code{django/django}, accounts for \SI{46.2}{\percent} of the Python benchmark
and resolves at \SI{76.6}{\percent}, just above the global mean. The lowest-rate
repository is \code{pylint-dev/pylint} at \SI{30.0}{\percent} ($n{=}10$). The
auxiliary per-year CSV remains available in the artifact; across
2017--2023, year-level resolution stays in the \SIrange{66}{94}{\percent} band, with the
high end driven by the small 2017 slice. We report the structural
repository pattern rather than mechanism because the workspace does not
contain a controlled study of repository-internal API churn or year
effects.

\begin{table}[t]
\centering
\scriptsize
\caption{SWE-bench Verified resolution by repository.}
\label{tab:byrepo}
\resizebox{\columnwidth}{!}{%
\begin{tabular}{@{}lrr||lrr@{}}
\toprule
Repository & Resolved/Total & Rate & Repository & Resolved/Total & Rate \\
\midrule
astropy/astropy & 13/22 & 59.1\% & pydata/xarray & 18/22 & 81.8\% \\
django/django & 177/231 & 76.6\% & pylint-dev/pylint & 3/10 & 30.0\% \\
matplotlib/matplotlib & 23/34 & 67.7\% & pytest-dev/pytest & 16/19 & 84.2\% \\
mwaskom/seaborn & 1/2 & 50.0\% & scikit-learn/scikit-learn & 27/32 & 84.4\% \\
pallets/flask & 1/1 & 100.0\% & sphinx-doc/sphinx & 30/44 & 68.2\% \\
psf/requests & 8/8 & 100.0\% & sympy/sympy & 57/75 & 76.0\% \\
\midrule
\multicolumn{6}{c}{\textbf{Overall: 374/500 (74.8\%)}} \\
\bottomrule
\end{tabular}%
}

\end{table}

\textbf{Threats.}
Python cross-system comparisons against published numbers from other
research groups use heterogeneous evaluation stacks; we therefore
restrict our Python headline number to the official evaluator and use
paired per-instance outcome vectors (rather than aggregate rates) for
all Python peer comparisons in RQ2. A run-to-run variance estimate beyond
the eight-run sweep would require additional eight-run sweeps under
the same Orchestra configuration; we instead report the per-run
pass@1 dispersion in RQ5 (eight independent decoding seeds,
$\sigma{=}3.12$ resolved instances, range 334--343/500), which we treat
as the within-configuration variance estimate for the headline.

\subsection{RQ2: Peer-Ranked Position With Multiplicity-Corrected Tests}
\label{subsec:rq2}

\textbf{Setup.}
We rank our system on a frozen snapshot of 135 publicly cataloged
SWE-bench Verified as of April~28, 2026, and run paired
McNemar exact tests~\cite{mcnemar1947note} on
the per-instance outcome vectors of a curated set of 17 open-weight and 7 closed-frontier peers.  Multiplicity is
controlled across the 24 paired tests with both Holm--Bonferroni
(family-wise error rate) and Benjamini--Hochberg (false discovery
rate) procedures~\cite{holm1979simple,benjamini1995controlling}, each with target $\alpha=0.05$ or $q=0.05$.

\textbf{Result.}
On the frozen Python leaderboard snapshot, the configuration ranks
\textbf{12th overall} and is the highest-ranked open-weight system;
the next open-weight submission, \emph{Lingxi v1.5 + Kimi-K2},
resolves 356/500 (\SI{71.2}{\percent}). Under Benjamini--Hochberg FDR control
($q\le0.05$), the configuration significantly outperforms 16 of 17
curated open-weight peers, including a 480-billion-parameter Qwen3-Coder
configuration ($\Delta=+26$ instances, BH-FDR $q=0.011$). The single
peer the configuration does not significantly outperform under
either FWER or FDR control is \emph{Lingxi v1.5 + Kimi-K2} (raw
$p=0.054$, $\Delta=+18$). Among closed-frontier peers, the
configuration is statistically indistinguishable at $\alpha=0.05$
from \emph{Sonar + Claude-Sonnet-4.5} ($\Delta=0$),
\emph{OpenHands + Claude-Opus-4.5}, \emph{live-SWE + Gemini-3-Pro},
and \emph{OpenHands + GPT-5}; significantly outperforms
\emph{OpenHands + Claude-4-Sonnet} (FDR-significant); and is
significantly behind the two top \emph{Sonar} / \emph{live-SWE +
Claude-Opus-4.5} builds (FDR-significant; $\Delta=-22$ in both
cases). The full pair-wise tables list raw McNemar $p$-values,
Holm-corrected $p$-values, BH-corrected $q$-values, and conditional
log-odds-ratio effect sizes with exact \SI{95}{\percent} Clopper--Pearson
confidence intervals~\cite{clopper1934use}.

\begin{figure}[t]
  \centering
  \includegraphics[width=\columnwidth]{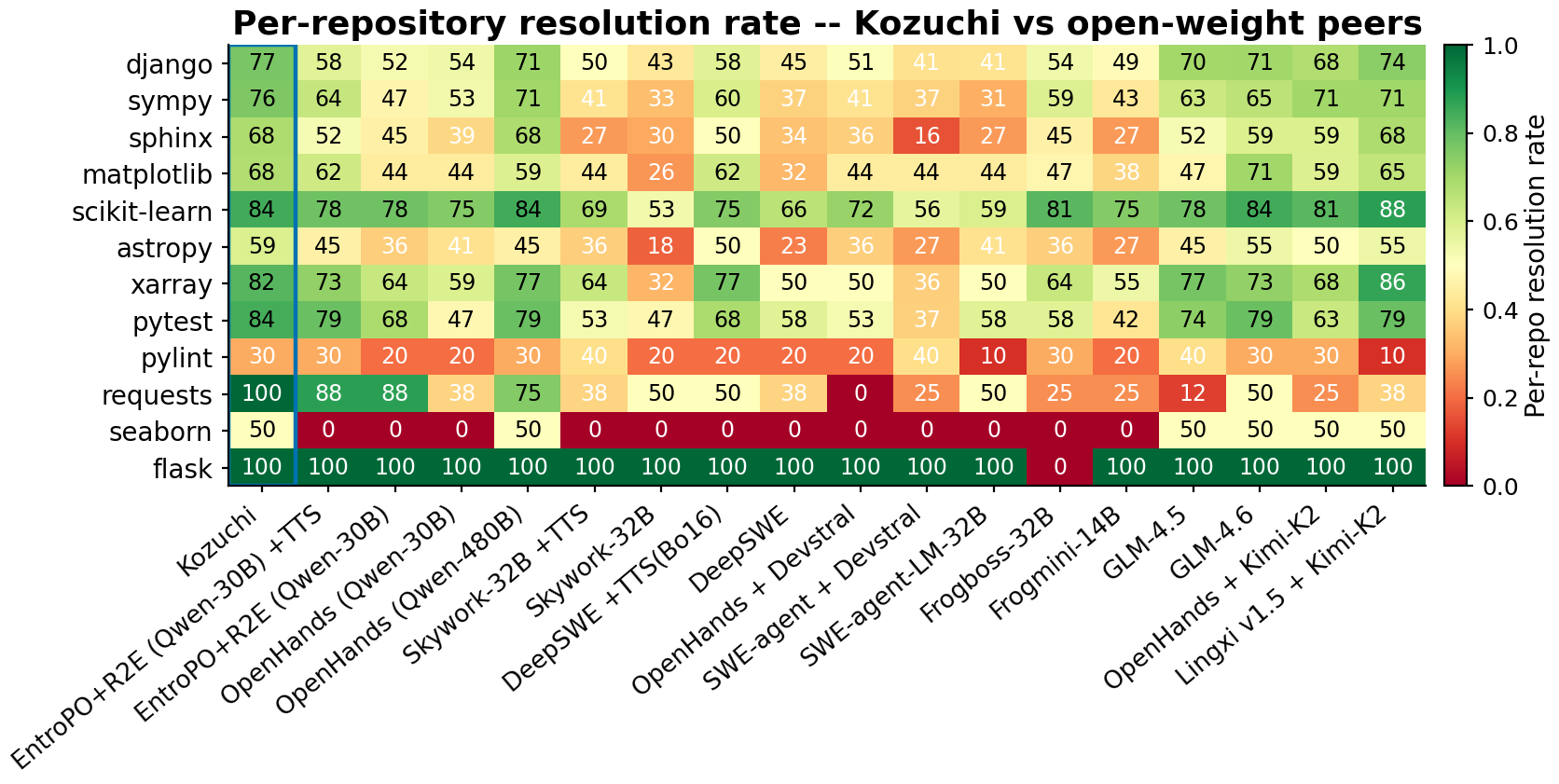}
  \caption{Per-repository resolution rate for Kozuchi agent (first column on the left) and the curated
  open-weight peers.}
  \Description{Heatmap comparing per-repository resolution rates for
  Kozuchi agent and open-weight peer agents across the 12 repositories in
  SWE-bench Verified. Kozuchi agent leads or ties the strongest peer on
  several high- and medium-volume repositories including django, sympy,
  sphinx, and pytest, while pylint remains low for most systems.}
  \label{fig:per-repo-peers}
\end{figure}

\textbf{Per-repository peer structure.}
Figure~\ref{fig:per-repo-peers} shows that the open-weight lead is not
only an aggregate leaderboard artifact. On the largest benchmark slice,
\code{django/django} ($n=231$), Kozuchi agent resolves \SI{76.6}{\percent}, ahead of the
next open-weight peer in the heatmap at \SI{74.0}{\percent}; on \code{sympy/sympy}
($n=75$) it resolves \SI{76.0}{\percent}, compared with \SI{70.7}{\percent} for the strongest
peers; and on \code{pytest-dev/pytest} ($n=19$) it reaches \SI{84.2}{\percent},
above the nearest peers at \SI{78.9}{\percent}. The same pattern appears on several
medium-size repositories: \code{sphinx-doc/sphinx} is tied for best at
\SI{68.2}{\percent}, and \code{pydata/xarray} is near the top at \SI{81.8}{\percent}. The main
exception is \code{pylint-dev/pylint}, where Kozuchi agent resolves only
\SI{30.0}{\percent} ($3/10$) and several peers reach \SI{40.0}{\percent}. We therefore interpret
the peer advantage as broad but not uniform: it is strongest on the
large repositories that dominate the Verified denominator, while
low-count repositories should be read as diagnostic failure clusters
rather than stable rank evidence. \S\ref{subsec:rq6} extends the peer
position analysis to Java, where the same 27-billion-parameter model configuration \textbf{ranks first}
among strict open-weight submissions.

\textbf{Threats.}
Ranks use a frozen snapshot of cataloged submissions.

\subsection{RQ3: Where Does Performance Vary?}
\label{subsec:rq3}

\textbf{Setup.}
For the Python SWE-bench Verified run, we stratify resolution by patch size, by per-instance LLM-call count,
and by phase-level events extracted from the 495 trajectory JSON
files; we also report a multivariate logistic regression with
cluster-robust standard errors (clustering at the repository level)
that adjusts trajectory features for external consensus hardness
proxies. Because the consensus terms use peer outcomes, we treat this
model as an explanatory diagnostic rather than as a deployable
predictor.

\textbf{Patch size.}
Resolution rate decreases monotonically with the line-of-code (LOC)
churn of the produced patch: from \SI{81.7}{\percent} at \SIrange{1}{4}{\loc} to \SI{41.7}{\percent} at
$>$\SI{100}{\loc}. The point-biserial correlation between LOC churn and
the binary resolution outcome is $r=-0.197$
($p=1.1{\times}10^{-5}$), the strongest single feature signal in
the dataset. The mean LOC added per patch is \SI{10.5}{\loc} (median
\SI{3}{\loc}) for resolved instances versus \SI{21.9}{\loc} (median
\SI{8}{\loc}) for unresolved instances.

\textbf{Trajectory effort.}
Trajectory-level effort metrics (per-instance LLM calls, runtime,
prompt tokens, bash invocations) all carry \emph{negative}
correlations with resolution. We interpret this signal as a
hardness-selection effect: harder instances cost more effort and are
also more likely to fail. We do not interpret it as ``effort hurts'';
no effort metric is positively correlated with resolution at
$\alpha=0.05$, which we report as a structural ceiling on what
further inference compute can buy at fixed scaffold and backbone.

\textbf{Phase-level behaviour.}
All 495 trajectories visit every one of the eight phases; rework is
concentrated in the \code{CODE\_FIX} $\leftrightarrow$ \code{VERIFY\_PATCH}
loop. \code{VERIFY\_PATCH} issues a \code{GIVEUP} in \SI{18.2}{\percent} of
trajectories (90/495, \SI{95}{\percent} CI [\SI{15.0}{\percent}, \SI{21.8}{\percent}]); \code{CODE\_FIX}
itself rolls back in \SI{1.0}{\percent} of cases; every other phase rolls back
zero times. Unresolved trajectories fire roughly \SI{28}{\percent} more
\code{CODE\_FIX} messages than resolved ones (mean 170 vs.\ 132)
and report 0.67 vs.\ 0.52 mean \code{VERIFY\_PATCH} \code{GIVEUP}s
per instance.

\textbf{Multivariate adjustment.}
A logistic regression of the binary resolution outcome on
log-transformed LLM calls, patch churn, runtime, and two external
consensus hardness proxies (\code{qwen\_consensus} and
\code{frontier\_consensus}) shows that the consensus terms dominate
the multivariate signal. After this adjustment, the trajectory-effort
and patch-churn coefficients are not significant at $\alpha=0.05$.
This supports the hardness-selection interpretation above and prevents
the univariate patch-churn correlation from being read as an
independent causal predictor.

\textbf{Threats.}
The consensus-adjusted model requires other systems' outcomes, and its
coefficients should not be read causally.

\subsection{RQ4: Failure Taxonomy and Operational Reliability}
\label{subsec:rq4}

\textbf{Setup.}
For the Python SWE-bench Verified run, we classify each unresolved instance into one of five mutually
exclusive failure categories from the SWE-bench harness reports.
We then report the operational profile of the 495 trajectories that
produced a harness report:
phase visit completeness, exit status, and patch-application
reliability.

\textbf{Failure taxonomy.}
Of 126 unresolved instances, 115 (\SI{91.3}{\percent}) are
\code{WRONG\_FIX}: the patch applies cleanly under the harness but
does not flip the hidden \code{FAIL\_TO\_PASS} tests.  6 (\SI{4.8}{\percent}) are
\code{REGRESSION}: the patch applies but breaks at least one
\code{PASS\_TO\_PASS} test. 5 (\SI{4.0}{\percent}) are
\code{MISSING\_ARTEFACT}: no trajectory or harness report was
persisted for the instance under the published bundle. The
\code{PATCH\_DID\_NOT\_APPLY} and \code{EMPTY\_PATCH} categories are
both 0.  The taxonomy is summarised in Table~\ref{tab:failuremodes}.

\begin{table}[t]
\centering
\scriptsize
\caption{Failure taxonomy for unresolved instances.}
\label{tab:failuremodes}
\begin{tabular}{@{}lrr@{}}
\toprule
Category & Count & Share \\
\midrule
\code{WRONG\_FIX}             & 115 & \SI{91.3}{\percent} \\
\code{REGRESSION}             &   6 &  \SI{4.8}{\percent} \\
\code{MISSING\_ARTEFACT}      &   5 &  \SI{4.0}{\percent} \\
\code{PATCH\_DID\_NOT\_APPLY} &   0 &  \SI{0.0}{\percent} \\
\code{EMPTY\_PATCH}           &   0 &  \SI{0.0}{\percent} \\
\midrule
Total                         & 126 & \SI{100.0}{\percent} \\
\bottomrule
\end{tabular}
\end{table}

\textbf{Operational reliability.}
All 495 trajectories report \code{exit\_status = "Submitted"}; the
agent does not crash mid-flight in this run. Of the 495 produced
patches, 494 apply cleanly through the SWE-bench harness, giving a
single edit-layer failure (\SI{0.20}{\percent}). Combined with the failure
taxonomy, this isolates the residual gap as a semantic-correctness
problem rather than an edit-format problem: the agent essentially
always emits a syntactically acceptable patch, but in \SI{91.3}{\percent} of
unresolved cases the patch is semantically wrong.

\textbf{Threats.}
\code{MISSING\_ARTEFACT} cases are operational rather than
modelling failures: under stricter persistence handling some of
them might be recovered at no algorithmic cost. They are not
removed from the Python headline denominator in this paper, but their
existence inflates the unresolved bucket by up to 5/500 (\SI{1.0}{\percentagepoint}).

\subsection{RQ5: TTS@8 Selection Lift and Cost--Diversity Envelope}
\label{subsec:rq5}

\textbf{Setup.}
For Python, we compare per-run pass@1 patches against the cross-agent
TTS@8 selection on the same eight runs. Cost is measured in additional
compute: $8\times$ candidate generation plus the cross-agent test
cross-application matrix. For Java, the shipped bundle exposes the
xcheck-selected source-run counts rather than complete per-leg pass@1
reports, so we use those counts only as a source-run audit of the
selected Java outcome.

\textbf{Per-run pass@1.}
The pass@1/pass@k terminology follows standard code-generation evaluation practice~\cite{chen2021codex}.
The eight Python pass@1 runs are tightly clustered: they resolve
334--343 of 500 instances (\SIrange{66.8}{68.6}{\percent}), with mean
338.6/500 = \SI{67.7}{\percent} and population standard deviation 3.12 resolved
instances (\SI{0.62}{\percentagepoint}). The submitted cross-agent selection raises this
average-run baseline to 374/500 = \SI{74.8}{\percent} on the official evaluator
and 376/500 = \SI{75.2}{\percent} on the internal Docker re-grade, a gain of
35.4--37.4 resolved instances (\SIrange{+7.1}{7.5}{\percentagepoint}) over mean pass@1.
For Java, the xcheck artifact selects 15.9 source-run candidates on
average and resolves 5.1 of them (population standard deviation
1.54), yielding 41/128 = \SI{32.03}{\percent} overall and a \SI{32.3}{\percent} conditional
resolved rate over selected source-run contributions.

\textbf{Selection lift.}
On Python, the cross-agent selector resolves 376/500 = \SI{75.2}{\percent} on the internal
Docker re-grade (Table~\ref{tab:xcheck}) and 374/500 = \SI{74.8}{\percent} on the
official evaluator. On 302 of 495 covered instances the
selector chooses a different candidate than the size-ordered first
candidate. The order baseline (first-run-wins) resolves 362/500 =
\SI{72.4}{\percent}; the selector therefore contributes +14 absolute (\SI{+2.8}{\percentagepoint})
over the order baseline and +33 to +42 absolute over individual
per-run pass@1.

\begin{table}[t]
\centering
\scriptsize
\caption{Artifact-based TTS@8 ablation. Scope: this table isolates
candidate count ($K$) and selector design only; phase graph, formatter,
persistent state, and tool sandbox are not ablated and are evidenced
operationally in \S\ref{sec:design} and \S\ref{sec:lessons}.}
\label{tab:selector-ablation}
\begin{tabular}{@{}p{0.34\columnwidth}rrp{0.27\columnwidth}@{}}
\toprule
Variant & Resolved & Rate & Note \\
\midrule
$K{=}1$, no selector (mean) & 338.6 & 67.7\% & eight-run mean \\
$K{=}1$, no selector (best) & 343 & 68.6\% & best leg \\
$K{=}8$, order baseline & 362 & 72.4\% & no cross-agent ranking \\
$K{=}8$, self-tests only & 346 & 69.2\% & estimated \\
$K{=}8$, cross-agent selector & 376 & 75.2\% & selected \\
$K{=}8$, oracle & 408 & 81.6\% & upper bound \\
\bottomrule
\end{tabular}

\end{table}

\textbf{Selector ablation.}
Table~\ref{tab:selector-ablation} separates the measured Python TTS@8 gain into
candidate generation and selection effects. Moving from a single
candidate to eight candidates creates the 408/500 oracle ceiling, while
the cross-agent selector captures 376 of those instances on the
internal Docker re-grade. A diagnostic self-tests-only selector,
computed from the same archived cross-agent selection tables by scoring
each candidate only on
its own generated tests and then checking the chosen run's per-run
report, resolves 346/500. The gap between self-tests-only selection
and the cross-agent selector's 376/500 result supports the specific
value of cross-agent test application, while the oracle gap leaves 32
internal Docker re-grade wins for better selection at fixed candidates.
This ablation isolates candidate count and selector signals; the other
harness mechanisms below are supported by operational signatures rather
than by controlled component-removal reruns.

\textbf{Diversity ceiling.}
Across the eight Python runs the union of resolved instances is 408 and
the intersection is 234. The selector's 376 corresponds to about
\SI{92}{\percent} of the internal Docker re-grade oracle ceiling ($376/408$). Pairwise Jaccard similarity
over \emph{completed} instances ranges from 0.943 to 0.965 (mean
0.957), so the diversity is in \emph{which} instances are resolved
rather than in coverage of the benchmark itself.

\textbf{Compute envelope.}
On Python, TTS@8 buys the +14 absolute lift over the size-order baseline at
$8\times$ candidate generation plus a separately logged selector
matrix: 21{,}635 patch--suite executions over 495 covered instances
(mean 5.59 deduplicated patches $\times$ 7.80 archived suites,
versus a naive $8\times8\times495=31{,}680$ matrix), booked as
50 cross-check shards in \code{xcheck\_manifest.json}. We measure
candidate-generation cost from the published trajectories rather than
from job logs.  Across the 495 trajectories, per-instance medians are
490 LLM calls, 6.20M prompt
tokens, 83{,}100 completion tokens, and \SI{0.77}{\hour} of wall-clock; the p95
tail extends to 1{,}099 LLM calls, 16.6M prompt tokens, and
\SI{1.94}{\hour}. Aggregated over the full eight-run candidate generation, the
configuration consumes approximately
$3.8{\times}10^{9}$ prompt tokens and
$5.8{\times}10^{7}$ completion tokens, so the workload is
prompt-token dominated, consistent with the persistent handover and
shared-state mechanism described in \S\ref{sec:design}.
Because the trajectory bundle records wall-clock but not
GPU-utilisation traces, we can bound rather than measure GPU cost:
mean per-instance wall-clock is \SI{0.95}{\hour} on the two-GPU tensor-parallel
serving configuration, so summing instance wall-clock across the full
eight-run generation gives an upper bound of roughly
$7.5{\times}10^{3}$ GPU-hours ($8 \times 495 \times 0.95\,\mathrm{h}
\times 2$). This is an upper bound because the vLLM servers batch many
concurrent instances on the same GPUs; per-instance wall-clock
double-counts shared occupancy. Accounted GPU-hours would require
cluster-side telemetry we did not collect (Table~\ref{tab:evidence-map}).
The compute--resolution Pareto curve over the 374 resolved instances
shows diminishing returns in the long tail: \SI{80}{\percent} of resolved instances
finish within 653 LLM calls per instance and \SI{95}{\percent} within 1{,}069, but
moving from the \SI{90}{\percent} point to the \SI{99}{\percent} point requires increasing the
budget from 820 to 1{,}547 calls. We report this Pareto as a descriptive
cost curve over \emph{this} run rather than as a prescription; an
early-stopping policy that exploits it would need its own controlled
study.

\textbf{Scaling under tighter budgets.}
For organizations without the infrastructure to spend the full TTS@8
envelope, the measured cost structure suggests three policies, none of
which we have validated with controlled runs: (i) a per-instance
call or token budget that stops the long tail (cutting at the \SI{90}{\percent}
Pareto point of 820 calls sacrifices at most the last decile of
resolved instances in this run); (ii) adaptive candidate count, since
234 of 500 instances are resolved by all eight runs and would be
selected correctly from far fewer candidates; and (iii) skipping the
cross-check matrix when deduplicated patches collapse to a single
candidate. Each policy needs its own controlled study before
deployment.

\textbf{Threats.}
Cross-agent testing requires that each candidate run produces
high-quality agent-generated tests; weak tests directly weaken the
selector's discrimination. A controlled comparison of direct versus
debate-style selection on the same trajectory bundle would require
a paired selector run with otherwise identical inputs. The compute
envelope is reported in LLM calls,
prompt/completion tokens, and wall-clock seconds, with GPU-hours given
only as a wall-clock-derived estimate, because the trajectory bundle
does not record GPU-utilisation traces; accounted GPU-hours require
cluster-side telemetry we did not collect. The ablation in
Table~\ref{tab:selector-ablation} isolates candidate count and selector
signals, but it does not isolate the causal contribution of the phase
graph, persistent handover, action formatter, tracing tools, or guarded
editing; those components require separate controlled reruns
(Table~\ref{tab:evidence-map}). The
auditable selector boundary is the archived suite outcome and patch
apply-status table, so hidden SWE-bench outcomes are unavailable until
after patch selection.

\subsection{RQ6: Cross-Language Transfer to Multi-SWE-bench Java}
\label{subsec:rq6}

\begin{figure}[t]
  \centering
  \includegraphics[width=\columnwidth]{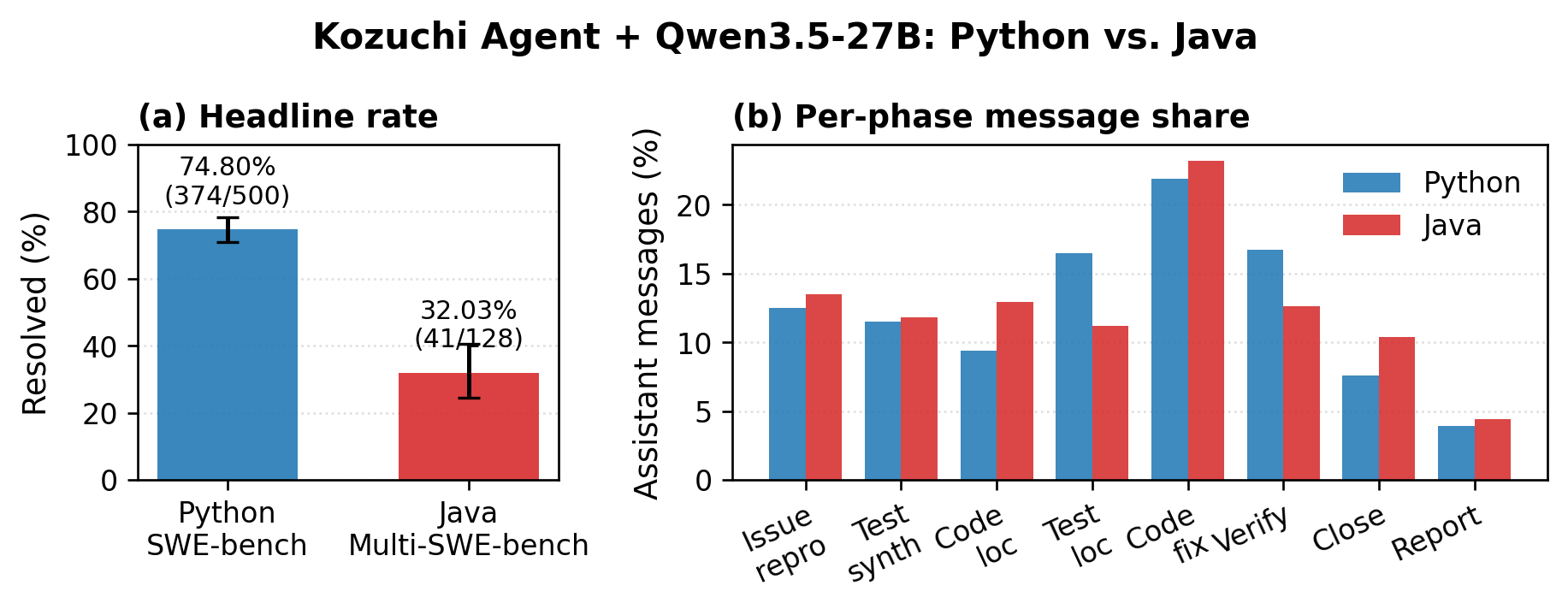}
  \caption{Cross-language transfer summary.}
  \Description{Two-panel comparison. The first panel shows Python at
  374 of 500 resolved and Java at 41 of 128 resolved with Wilson
  confidence intervals. The second panel overlays the eight phase message
  shares for the Python and Java trajectories, showing similar shape
  across the phase graph.}
  \label{fig:cross-track}
\end{figure}

We re-ran the same Qwen3.5-27B agent, eight-phase scaffold, and $K{=}8$
selection pattern on Multi-SWE-bench Java.  Only the corpus and
harness change: Java uses Maven/Gradle and strict xcheck@8, while the
Python headline uses pytest and TTS@8 (\S\ref{sec:design-tts}).  The
phase graph, action contract, persistent state, and selection family
remain the same. Figure~\ref{fig:cross-track} shows headline resolution
with Wilson intervals~\cite{wilson1927score} and per-phase
assistant-message share, while Table~\ref{tab:cross-track} summarizes
the matched Python--Java evidence.

\begin{table}[t]
\centering
\scriptsize
\caption{Python--Java cross-track evidence for the same 27-billion-parameter Kozuchi agent configuration.}
\label{tab:cross-track}
\begin{tabular}{@{}p{0.37\columnwidth}p{0.28\columnwidth}p{0.27\columnwidth}@{}}
\toprule
Signal & Python & Java \\
\midrule
Benchmark & SWE-bench Verified ($n{=}500$) & Multi-SWE-bench Java ($n{=}128$) \\
Resolved (Wilson 95\% CI) & 374/500 = 74.80\% [70.82, 78.41] & 41/128 = 32.03\% [24.57, 40.54] \\
Leaderboard position & 12/135 overall & 4/42 overall \\
Open-weight rank & 1st curated open-weight & 1st strict open-weight \\
BH-FDR peer wins & 16/17 open-weight peers & 34/41 Java peers \\
Same-family open margin & +26 vs. Qwen3-Coder-480B & +38 vs. Qwen2.5-72B \\
Rework-loop phase delta & baseline & CODE\_FIX +1.3 pp; VERIFY\_PATCH -4.1 pp \\
Median LLM calls & 490 & 529 \\
Nonempty applying patch output & 494/495 = 99.80\% & 124/128 = 96.9\% \\
Globally novel solves & 1 & 1 \\
\bottomrule
\end{tabular}

\end{table}

The Java run resolves 41/128 instances (\SI{32.03}{\percent}, Wilson \SI{95}{\percent} CI
[\SI{24.57}{\percent}, \SI{40.54}{\percent}]), ranks 4 of 42 overall, and is first among strict
open-weight submissions. Under BH-FDR control ($q\le0.05$), it
significantly outperforms 34 of 41 Java peers, including all
72-billion-parameter
Qwen2.5 open-weight builds ($\Delta=+27$ to $+38$). The only
FDR-significant adverse comparison is \emph{CodeArts-Agent +
CodeArts-MiniMax-M2.5} ($\Delta=-15$, $q=0.011$), with a much larger
closed-weight backbone.

The scaffold behaviour transfers more strongly than the absolute score:
per-phase message share is within $\pm$\SI{5}{\percentagepoint} on every phase, rework still
centres on \code{CODE\_FIX}$\leftrightarrow$\code{VERIFY\_PATCH}, and six
of seven effort medians sit within $\pm\SI{20}{\percent}$ of Python parity. Thus the
\SI{43}{\percentagepoint} headline gap is not a compute gap. The Java failure profile is
instead language/harness-specific: \code{no\_fix\_test\_results} (\SI{25}{\percent}),
\code{report\_valid\_not\_leaderboard\_resolved} (\SI{9}{\percent}), and
\code{anomalous\_test\_pattern} (\SI{6}{\percent}) do not appear on Python; edit-layer
reliability is lower (\SI{96.9}{\percent} vs. \SI{99.8}{\percent}); and \code{elastic/logstash}
accounts for 38/128 Java instances, with 0/38 solved by Kozuchi agent and
32/38 unresolved by every catalogued peer. We therefore claim
language-agnostic harness transfer, not equal language fluency.

The Java leaderboard mixes open-weight, semi-open, undeclared, and
closed-frontier peers; our open-weight claim uses the strict
``decision-LLM weights publicly released'' subset. As in RQ5, the Java
run does not separately ablate scaffold, and selector
contributions.
\section{Operational Planning and Lessons Learned}
\label{sec:lessons}

The operational lessons from benchmark-scale use are practical rather
than architectural.  First,
phase boundaries kept long-horizon runs structured and auditable: every Python
TTS@8 trajectory visits all eight phases, rework concentrates in
\code{CODE\_FIX}$\leftrightarrow$\code{VERIFY\_PATCH}, and the same phase
shape transfers to Java within $\pm5$ percentage points per phase
(\S\ref{subsec:rq6}).  Second, model-family variation is cheapest to
handle at the action-format boundary: switching tool-call syntax is a
configuration change rather than an agent fork.  Third, CI reuse is a
core evaluation lever, because operators can hold trajectories,
selection bundles, or environments fixed while changing one stage.  The
same principle applies to filesystem-backed state: reproduction notes,
generated tests, traces, and handover memos survive context compression
and make runs reviewable after the fact.  Finally, cross-agent testing
turns candidate diversity into accuracy without a learned verifier, while
one cluster abstraction keeps multi-site runs reviewable.

The artifact-derived operational indicators are consistent with those
lessons: in the inventoried workflow, five operator touch-points become
one CI push; the about-70-engineer-minute-per-cycle figure is an estimate
derived from the published per-stage inventory and the SWE-bench
Docker-runtime / CI-adoption anchors in prior work~\cite{epoch2024swebenchdocker,hilton2016ci}.
It is not a measured before/after from a controlled developer pilot, and
we report it as an order-of-magnitude operational claim rather than a
productivity metric. The runtime exposes four action formats across 17
model configs, six of nine CI stages are reusable, and five cluster
environments sit behind \code{ENV\_NAME}.  On Python, 494/495 produced
patches apply cleanly and the cloud-vs-Docker score drift is only 374
vs. 376 out of 500.  On Java, the same runtime contract reaches 41/128
(\SI{32.03}{\percent}) and ranks first among strict open-weight submissions, making
cross-language transfer a measured operational outcome rather than an
anecdotal consistency check.

\section{Conclusion}
\label{sec:conclusion}

\textbf{Operational readiness evidence.}
Kozuchi Agent targets conservative pre-production evaluation of
issue-to-patch agents on large repositories, with evidence from
SWE-bench (Python) and Multi-SWE-bench Java. Using a
locally hosted open-weight model, no fine-tuning,
eight candidate runs, and cross-agent testing, it resolves 374/500
SWE-bench Verified instances (\SI{74.80}{\percent}, Wilson \SI{95}{\percent} CI [\SI{70.82}{\percent},
\SI{78.41}{\percent}]; repository-clustered bootstrap [\SI{67.0}{\percent}, \SI{79.8}{\percent}]) and 41/128
Multi-SWE-bench Java instances (\SI{32.03}{\percent}) under the same agent,
backbone, and scaffold. It ranks 12th among 135 catalogued Python
submissions, is the highest-ranked open-weight Python system, ranks first
among strict open-weight Java, and recovers \SI{80}{\percent} of resolved
Python instances within 653 LLM calls.

\textbf{What we learned.}
The results point to harness engineering---phase decomposition,
single-command grammar, persistent shared filesystem, deterministic SE
tools, cross-agent test selection, cluster indirection, and CI reuse---
while ablations isolate only candidate count and selector signals; we
leave controlled per-mechanism ablations (no-phase, no-formatter,
no-Orchestra, no-tools) to a follow-up that can re-use the published
trajectory bundle without re-running TTS@8. The
phase decomposition, action grammar, persistent state, and cross-agent
selector transfer from Python to Java unchanged, supporting the narrower
cross-language harness-transfer claim. The selector adds +14 resolved Python instances;
494/495 Python patches apply cleanly; and remaining Python errors are
mainly semantic or selection failures: \SI{91.3}{\percent} of unresolved instances
have clean-applying patches that miss hidden tests, and 34 instances are
solved by some run but not selected.

\textbf{Next.}
Future work should ablate tools, selector design, and post-training; use
conversation audits for content-aware selection; connect failure-pattern
detection to harness updates; and use the compute--resolution Pareto
curve for controlled early stopping and adaptive candidate counts
(\S\ref{subsec:rq5}).
Industrial-impact claims are deliberately scoped to internal
evaluation-pipeline replacement (Table~\ref{tab:workflow-replacement},
\S\ref{sec:lessons}), cross-evaluator consistency (374/376 agreement),
and Python--Java benchmark transfer of the same agent. Production
deployment against proprietary repositories, end-user developer studies,
field A/B telemetry, and broader multilingual benchmarks are explicit
future work.

\section*{Contributions of Each Team}
\label{sec:contributions}

\paragraph{Fujitsu Research, Japan.}
\begin{itemize}
  \item Led the overall design and implementation of Kozuchi Agent.
  \item Conducted evaluations on SWE-bench.
\end{itemize}

\paragraph{Fujitsu Research of America, USA}
\begin{itemize}
  \item Led the writing of this paper.
  \item Designed the Verifier.
  \item Developed and prepared tools for Java.
  \item Conducted evaluations on Multi-SWE-bench Java.
\end{itemize}

\paragraph{Fujitsu Research of Europe, UK}
\begin{itemize}
  \item Developed the Software Engineering Tool Suite.
\end{itemize}

\paragraph{Fujitsu Research \& Development, China.}
\begin{itemize}
  \item Conducted evaluations of test-time selection.
\end{itemize}

\section*{Data Availability Statement}
\label{sec:artifact}
\label{sec:data}
The public review artifact, including all analysis, is available at
\url{https://github.com/marscod/kozuchi-agent-artifact} and Kozuchi Agent source code at: \url{https://github.com/FujitsuResearch/kozuchi-mini-swe-agent}.

The complete agent execution trajectories are available in the Zenodo
archival record (DOI: 10.5281/zenodo.19941649), together with the agent
source code, submission folders, logs, predictions, selector bundle,
generated summaries, figures, audits, and scripts needed to regenerate
RQ1--RQ6.

\nocite{qwen35,devstral,nemotron}
\balance
\bibliographystyle{ACM-Reference-Format}
\bibliography{references}

\end{document}